# ФОРМИРОВАНИЕ ТОНКИХ БУФЕРНЫХ СЛОЕВ GaAs НА ПОВЕРХНОСТИ КРЕМНИЯ ДЛЯ СВЕТОИЗЛУЧАЮЩИХ ПРИБОРОВ

©2024 г. В. В. Лендяшова[a, b, *], И. В. Илькив[a, b, **], Б. Р. Бородин[c], Д. А. Кириленко[c], А. С. Драгунова[d, b], Т. Шугабаев[a, b], Г. Э. Цырлин[a, b, e]

[a]*Санкт-Петербургский государственный университет, Санкт-Петербург, 199034 Россия*

[b]*Санкт-Петербургский национальный исследовательский академический университет им. Ж.И. Алферова РАН, Санкт-Петербург, 194021 Россия*

[c]*Физико-технический институт им. А.Ф. Иоффе РАН, Санкт-Петербург, 194021 Россия*

[d]*Международная лаборатория квантовой оптоэлектроники, Национальный исследовательский университет "Высшая школа экономики", Санкт-Петербург, 190008 Россия*

[e]*Университет ИТМО, Санкт-Петербург, 197101 Россия*

*\*e-mail: erilerican@gmail.com*

*\*\*e-mail: fiskerr@ymail.com*

В работе представлены экспериментальные результаты по исследованию процессов роста GaAs слоев на подложках кремния методом молекулярно-пучковой эпитаксии. Установлено, что формирование буферного Si слоя в едином ростовом процессе позволяет существенно повысить кристаллическое качество формируемых на его поверхности GaAs слоев, а также предотвратить формирование антифазных доменов как на разориентированных в направлении [110], так и на сингулярных на Si(100) подложках. Продемонстрировано, что применение циклического термического отжига при температурах 350-660°C в потоке атомов мышьяка позволяет снизить количество прорастающих дислокаций и повысить гладкость поверхности в GaAs слоев. При этом в статье рассматриваются возможные механизмы, приводящие к улучшению качества приповерхностных слоев GaAs. Показано, что полученные таким образом GaAs слои субмикронной толщины на сингулярных подложках Si(100) обладают среднеквадратичным значением шероховатости поверхности 1.9 нм. Представлена принципиальная возможность использования тонких GaAs слои на кремнии в качестве темплейтов для формирования на их основе светоизлучающих полупроводниковых гетероструктур с активной областью на основе самоорганизующихся InAs квантовых точек и InGaAs квантовой ямы. Показано, что они демонстрируют фотолюминесценцию в области длины волны излучения 1.2 мкм при комнатной температуре.

**Ключевые слова:** молекулярно-пучковая эпитаксия, полупроводники, кремний, арсенид галия, арсенид индия, арсенид индия галлия, субмикронные слои, квантовые точки, полупроводниковые гетероструктуры, телекоммуникации.

ВВЕДЕНИЕ

Интеграция полупроводниковых приборов на основе материалов типа $A^{III}B^{V}$ и кремния до сих пор представляет значительный интерес. Это открывает широкие возможности для создания активных приборов фотоники, таких как лазеры, светодиоды, фотодетекторы в области длин волн второго и третьего телекоммуникационных окон, объединенных напрямую с пассивными элементами – волноводами, фильтрами, разветвителями и сумматорами на базе кремниевой платформы [1–3]. Тем не менее, такая интеграция до сих пор остается достаточно сложной научной задачей. Один из перспективных подходов к ее решению основан на прямом синтезе материалов типа $A^{III}B^{V}$ на поверхности кремниевых подложек. Несмотря на то, что к настоящему времени были достигнуты определенные успехи в области получения планарных гетероструктур типа $A^{III}B^{V}$ на кремнии, а также создании на их основе инжекционных лазеров с непрерывным режимом работы при комнатной температуре [4], синтез буферных слоев с высоким кристаллическим качеством до сих пор представляет большую сложность. Это связано с тем, что наличие полярных плоскостей соединений типа $A^{III}B^{V}$ приводит к возникновению антифазных границ в процессе роста на Si поверхности, а несоответствие параметров решеток между ними – к появлению большого количества пронизывающих дислокаций [5]. Улучшение кристаллического качества формируемых на кремнии слоев типа $A^{III}B^{V}$ обычно достигается формированием достаточно толстых (порядка 5–7 мкм) буферных слоев градиентного состава [6,7], в том числе включающих напряженные слои сверхрешетки [8, 9], а также использованием подложек Si(100) с отклонением 4°–6° в направлении [110] [8], пространственно-ограниченном ростом [10, 11] и т.д. Однако, несмотря на достигнутые результаты [6–11], высокий расход материалов и технологическая сложность процесса формирования слоев до сих пор остаются основными проблемами подобных подходов. Поэтому все более актуальными становятся исследования, направленные, главным образом, на уменьшение толщины буферного слоя. Кроме того, высокий интерес вызывает использование вместо квантовых ям массивов самоорганизующихся квантовых точек (КТ) материалов типа $A^{III}B^{V}$, которые в меньшей степени подвержены влиянию дислокаций [12–20], в качестве активной области для создания светоизлучающих приборов.

Настоящая работа посвящена исследованию возможностей получения тонких эпитаксиальных слоев GaAs на поверхности подложек Si(100) методом молекулярно-пучковой эпитаксии (МПЭ), а также формирования на их основе светоизлучающих полупроводниковых гетероструктур.

## МЕТОДИКА ЭКСПЕРИМЕНТА

В качестве подложек были использованы круглые пластины Si(100) диаметром 50.8 мм как сингулярные Si(100), так и с разориентацией 4° в направлении [110]. Перед проведением экспериментов по росту осуществляли жидкофазное химическое травление подложек по методике, основанной на методе Шираки [21]. Синтез образцов осуществляли в едином технологическом цикле с использованием МПЭ-установки Riber Compact 21 EB200, оборудованной эффузионными источниками для роста соединений типа $A^{III}B^{V}$, а также электронно-лучевым испарителем для осаждения кремния. После формирования буферного слоя Si толщиной 50 нм осуществляли рост слоя GaAs общей толщиной 850 нм. Сначала был выращен низкотемпературный ($T_s = 350°C$) слой толщиной 150 нм, за которым следовал слой, полученный при промежуточной температуре ($T_s = 450°C$), толщиной 200 нм, включающий упруго напряженный фильтр-слой InGaAs толщиной 100 нм, и высокотемпературный ($T_s = 550°C$) слой толщиной 0.5 мкм.

Исследования морфологических особенностей синтезированных образцов осуществляли методами атомно-силовой микроскопии (АСМ) и растровой электронной микроскопии (РЭМ). Для этого были использован атомно-силовой микроскоп Ntegra Aura, работающий в полуконтактном режиме с использованием кремниевых зондов (HANC, TipsNano) с радиусом кривизны кончика <10 нм. Измерения оптических свойств проводили методом фотолюминесценции (ФЛ) при комнатной температуре. Накачку осуществляли зеленым лазером с длиной волны 527 нм, работающем на генерации второй гармоники. Спектры фотолюминесценции регистрировали с помощью InGaAs-фотодетектора. Ширина входной и выходной щелей монохроматора составляла 1500 мкм с решеткой 400 (1200).

## РЕЗУЛЬТАТЫ И ИХ ОБСУЖДЕНИЕ

Как уже было сказано, использование Si(100) подложек, разориентированных в направлении [110] на 4° и более градусов позволяет избежать образование антифазных границ в слоях GaAs за счет поверхности с двухатомным порядком ступеней. Однако для создания микроэлектронных приборов требуется использование сингулярных подложек кремния с разориентацией менее 0.5°. В свою очередь, двухатомный порядок ступеней может быть достигнут и на подложках разориентированных менее чем на 4°, в том числе сингулярных, как было недавно показано в работах [22, 23].

С этой целью был сформирован тонкий буферный Si слой с помощью описанного выше метода. Для этого предварительно подготовленные разориентированные и сингулярные пластины кремния загружали в МПЭ-установку, и проводили их термический отжиг при температуре 1100°C с целью удаления тонкого слоя оксида. Затем при температуре 600°C выращивали буферный слой Si толщиной 50 нм и осуществляли его высокотемпературный отжиг при 1100°C для достижения режима формирования эшелонов ступеней на сингулярной поверхности. После этого температуру подложки понижали до 350°C для роста буферного слоя GaAs. Далее выращивали низкотемпературный, зародышевый слой GaAs толщиной 150 нм, который является важным этапом формирования всех последующих слоев для подавления антифазных границ и уменьшения плотности прорастающих дислокаций [24]. Затем температуру подложки поднимали до 450°C и осуществляли последовательный рост 50 нм слоя GaAs, 100 нм слоя $In_{0.1}Ga_{0.9}As$ и 50 нм слоя GaAs. Использование тройного раствора в качестве фильтр-слоя также позволяет повысить вероятность аннигиляции пронизывающих дислокаций. После этого, при температуре 550°C выращивали слой GaAs толщиной 0.5 мкм и осуществляли циклический термический отжиг выращенных слоев, что, в свою очередь, также позволяет снизить плотность пронизывающих дислокаций [25].

На рис. 1а, 1в, 1д представлены РЭМ-изображения поверхности слоев GaAs, синтезированных на различных подложках. Видно, что на поверхности слоя GaAs, выращенного на сингулярной подложке Si(100) без буферного слоя кремния имеются множественные дефекты, включая ямки и их эшелоны. В свою очередь, на поверхности слоев GaAs, выращенных на подложках с тонким буферным Si слоем, подобные дефекты отсутствуют (рис. 1в, 1д). Следует отметить, что между слоем GaAs, выращенным на разориентированной подложке с использование буферного слоя кремния, и аналогичным слоем, выращенным на сингулярной подложке, существует заметное различие в морфологии поверхности. Слой GaAs, выращенный на Si(100), более гладкий и однородный. Более детальное исследование поверхности, выполненное с помощью АСМ, показало, что для слоев на разориентированных подложках Si среднеквадратичная шероховатость (RMS) в области сканирования 10×10 мкм составляет 3–5 нм (рис. 1б). В то же самое время для слоев, выращенных на Si(100) (рис. 1г), в области сканирования достигается среднеквадратичная шероховатость порядка 1.9 нм. Близкие значения среднеквадратичной шероховатости буферных слоев (например, около 2.85 нм [17]) в системе GaAs/Si были получены ранее только для случаев толстых буферов (более 2 мкм). Кроме того, важно отметить, что в обоих случаях отсутствуют границы антифазных доменов на поверхности слоев GaAs, что подтверждает достижение режима формирования эшелонов ступеней при росте на подложке Si(100). Таким образом, был разработан новый подход к формированию гладких слоев GaAs субмикронной толщины (850 нм).

Для оценки возможности использования GaAs слоев на кремнии в качестве темплейтов для последующего формирования на их основе светоизлучающих приборов были проведены дополнительные ростовые эксперименты. Для этого пластина кремния с буферным GaAs слоем была разделена на четыре одинаковые части, после чего 1/4 часть подложки Si(100) повторно загружали в МПЭ-установку для формирования слоев активной области. Был выполнен последовательный рост 150 нм слоя GaAs, 150 нм слоя $Al_{0.3}Ga_{0.7}As$ и 100 нм слоя GaAs при температуре 550°C. Затем при температуре 480°C осуществляли самоорганизующийся рост КТ InAs с эффективной толщиной осажденного материала 7.6 Å (2.5 монослоя) и покрывающей их квантовой ямы $In_{0.1}Ga_{0.9}As$ толщиной 5 нм (dots-under-the-well, DUWELL-структур [26]). После этого температуру подложки повышали до 590°C и выращивали 100 нм слой GaAs, поверх которого при температуре 550°C формировали двойной покровный слой из 100 нм $Al_{0.3}Ga_{0.7}As$ и 10 нм GaAs. По завершении роста образец охлаждали до комнатной температуры и выгружался из МПЭ-установки для исследования его оптических свойств.

На рис. 2 представлен спектр фотолюминесценции полученного образца. Максимум интенсивности фотолюминесценции находится около 1196 нм, а полная ширина на половине высоты составляет порядка 81 мэВ. Относительно широкий спектр может быть связан с близким к бимодальному распределением квантовых точек по размеру [27, 28]. Неоднородность размеров островков при относительно малых эффективных толщинах осажденного материала встречается в системах материалов InAs/InP, InAs/GaAs и может быть связана с наличием энергетических барьеров [29, 30]. Оптические свойства КТ InAs в GaAs на подложке кремния сравнивали с аналогичной светоизлучающей полупроводниковой гетероструктурой InAs/InGaAs на подложке GaAs. В результате было установлено, что интенсивность сигнала фотолюминесценции (в точке максимума) от созданной в настоящей работе DUWELL-структуры InAs/InGaAs в GaAs на подложке кремния, в сравнении с аналогичной гетероструктурой на подложке GaAs, составляет порядка 55%.

ЗАКЛЮЧЕНИЕ

В работе исследовано получение эпитаксиальных GaAs слоев субмикронной толщины на подложках кремния методом молекулярно-пучковой эпитаксии для создания светоизлучающих приборов. Было установлено, что использование тонких буферных Si слоев позволяет существенным образом повысить кристаллическое качество впоследствии формируемых слоев GaAs. На поверхности сформированных буферных слоев GaAs субмикронной толщины были синтезированы полупроводниковые гетероструктуры с активными слоями на основе КТ InAs и покрывающей их InGaAs квантовой ямы, демонстрирующие фотолюминесценцию в области длины волны излучения 1.2 мкм при комнатной температуре. Полученные результаты представляют значительный интерес для создания новых светоизлучающих устройств, монолитно объединенных с приборами на основе кремния.

**Конфликт интересов**: Авторы заявляют, что у них нет конфликта интересов.

Настоящая статья не содержит каких-либо исследований с участием людей в качестве объектов исследований.

# Formation of Thin GaAs Buffer Layers on Silicon for Light-Emitting Devices

V. V. Lendyashova[1,2,*], I. V. Ilkiv[1,2,**], B. R. Borodin[3], D. A. Kirilenko[3], A. S. Dragunova[4,2], T. Shugabaev[1,2], G. E. Cirlin[1,2,5]

[1] *Saint Petersburg State University, St.-Petersburg, 199034 Russia*

[2] *Alferov University, St.-Petersburg, 194021 Russia*

[3] *Ioffe Institute, St.-Petersburg, 194021 Russia*

[4] *International laboratory of quantum optoelectronics, HSE University, St.-Petersburg, 190008 Russia*

[5] *ITMO University, St.-Petersburg, 197101 Russia*

*e-mail: erilerican@gmail.com

**e-mail: fiskerr@ymail.com

This paper presents the experimental results on research of growth processes of GaAs layers on silicon substrates by molecular beam epitaxy. The formation of buffer Si layer in a single growth process has been found to significantly improve the crystalline quality of the GaAs layers formed on its surface, as well as to prevent the formation of anti-phase domains both on offcutted towards the [110] direction and on singular Si(100) substrates. It has been demonstrated that the use of cyclic thermal annealing at temperatures 350-660°C in the flow of arsenic atoms makes it possible to reduce the number of threading dislocations and increase the smoothness of the GaAs layers surface. At the same time, the article considers possible mechanisms that lead to an improvement in the quality of the surface layers of GaAs. It is shown that the thus obtained GaAs layers of submicron thickness on the singular Si(100) substrates have a mean square value of surface roughness 1.9 nm. The principal possibility of using thin GaAs layers on silicon as templates for forming on them light-emitting semiconductor heterostructures with active area based on self-organizing InAs quantum dots and InGaAs quantum well is presented. They are shown to exhibit photoluminescence at 1.2 μm at room temperature.

**Keywords:** molecular beam epitaxy, semiconductors, silicon, gallium arsenide, indium arsenide, indium gallium arsenide, submicron layers, quantum dots, semiconductor heterostructures, telecommunications.

ПОДПИСИ К РИСУНКАМ

**Рис.1** (а) РЭМ-изображение поверхности слоя GaAs, выращенного без буферного слоя Si и отжига на подложке Si(100); (б) АСМ-изображение и (в) РЭМ-изображение поверхности слоя GaAs, выращенного с буферным слоем Si на подложке Si(100) с разориентацией 4°; (в) АСМ-изображение и (г) РЭМ-изображение поверхности слоя GaAs, выращенного с буферным слоем Si на подложке Si(100). Для АСМ-изображений область сканирования составляет 10×10 мкм.

**Рис.2** Спектр фотолюминесценции от InAs квантовых точек, покрытых квантовой ямой InGaAs, внедренных в слой GaAs на подложке Si(100).

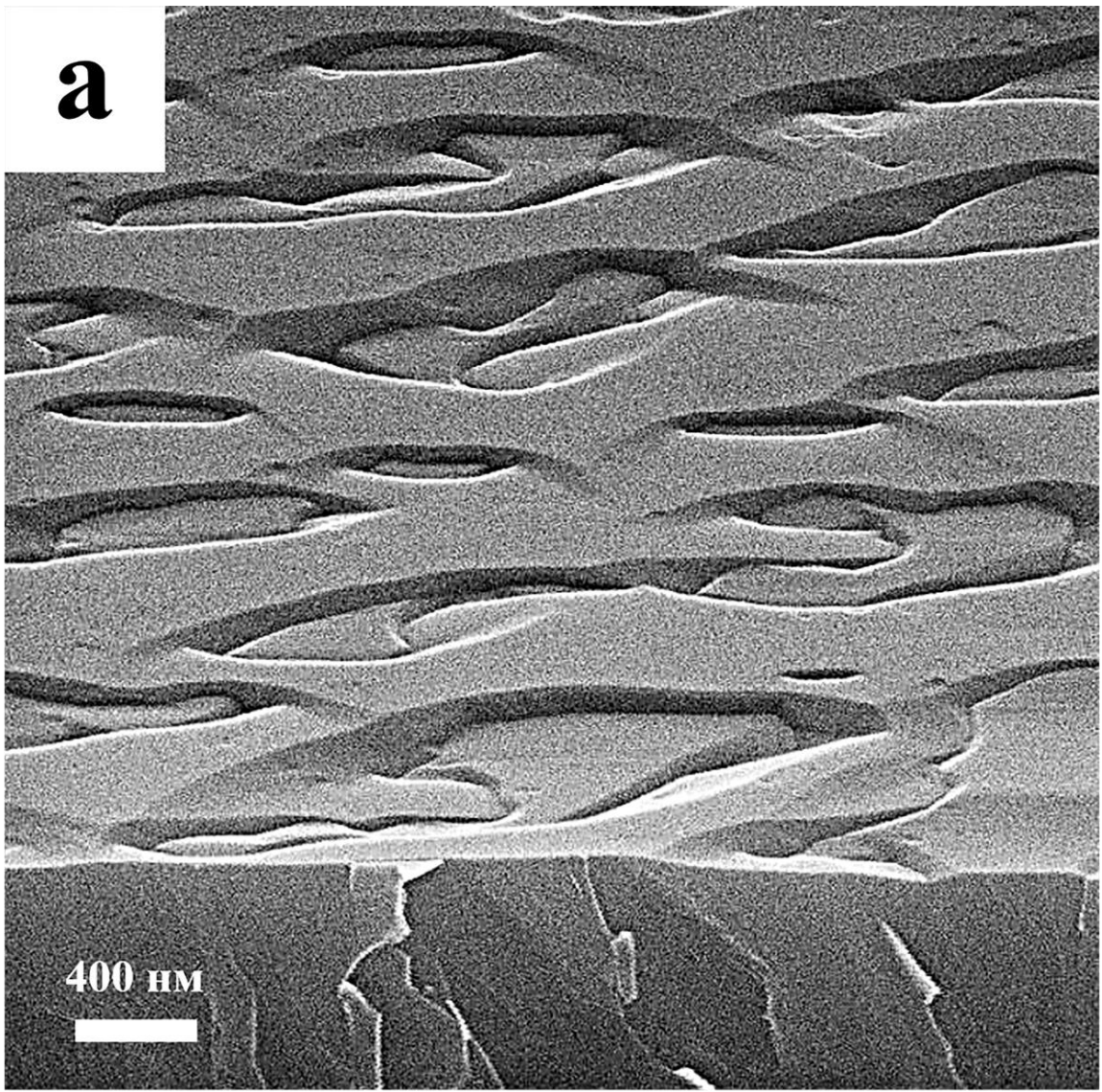

(б)

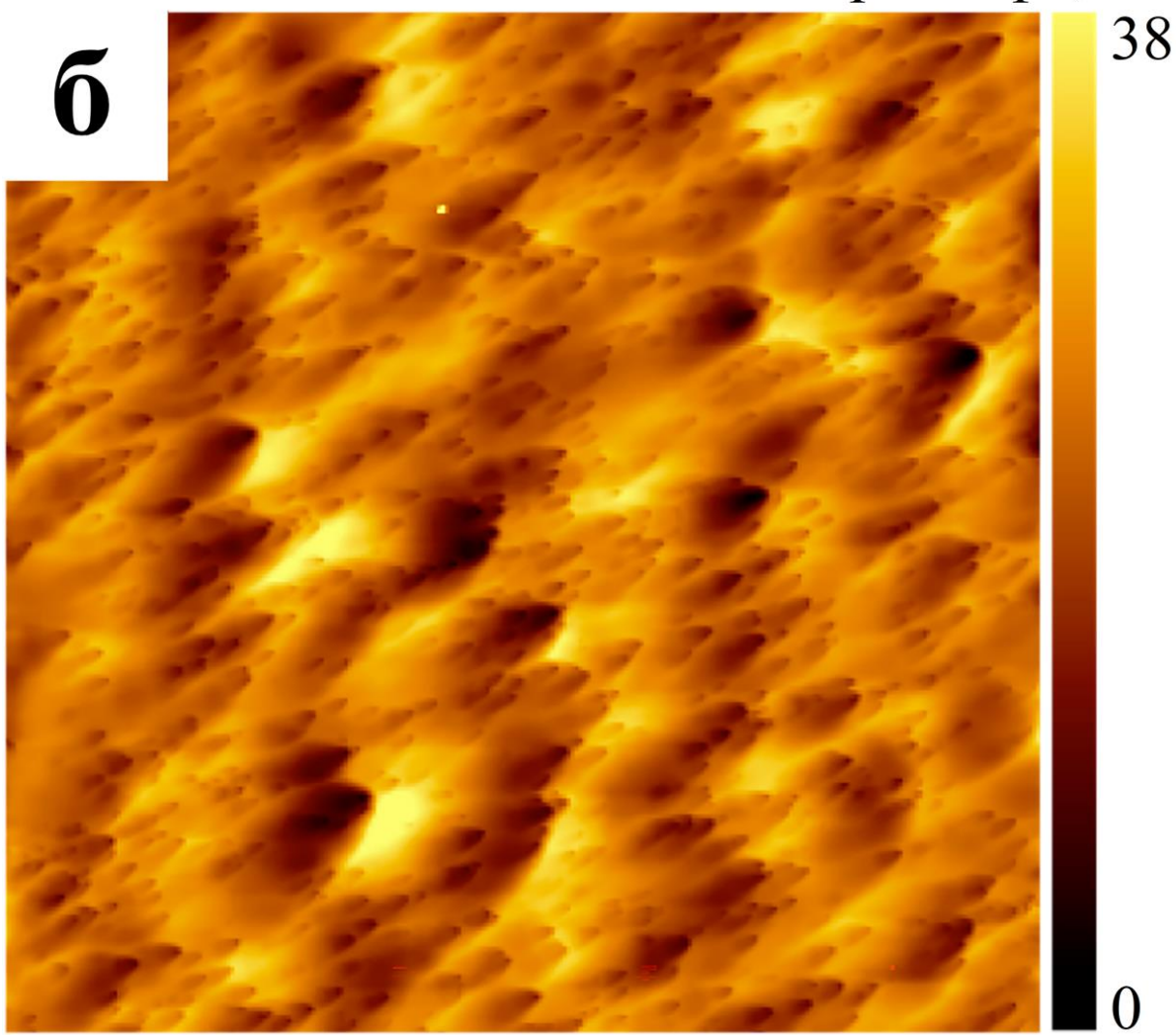

(в)

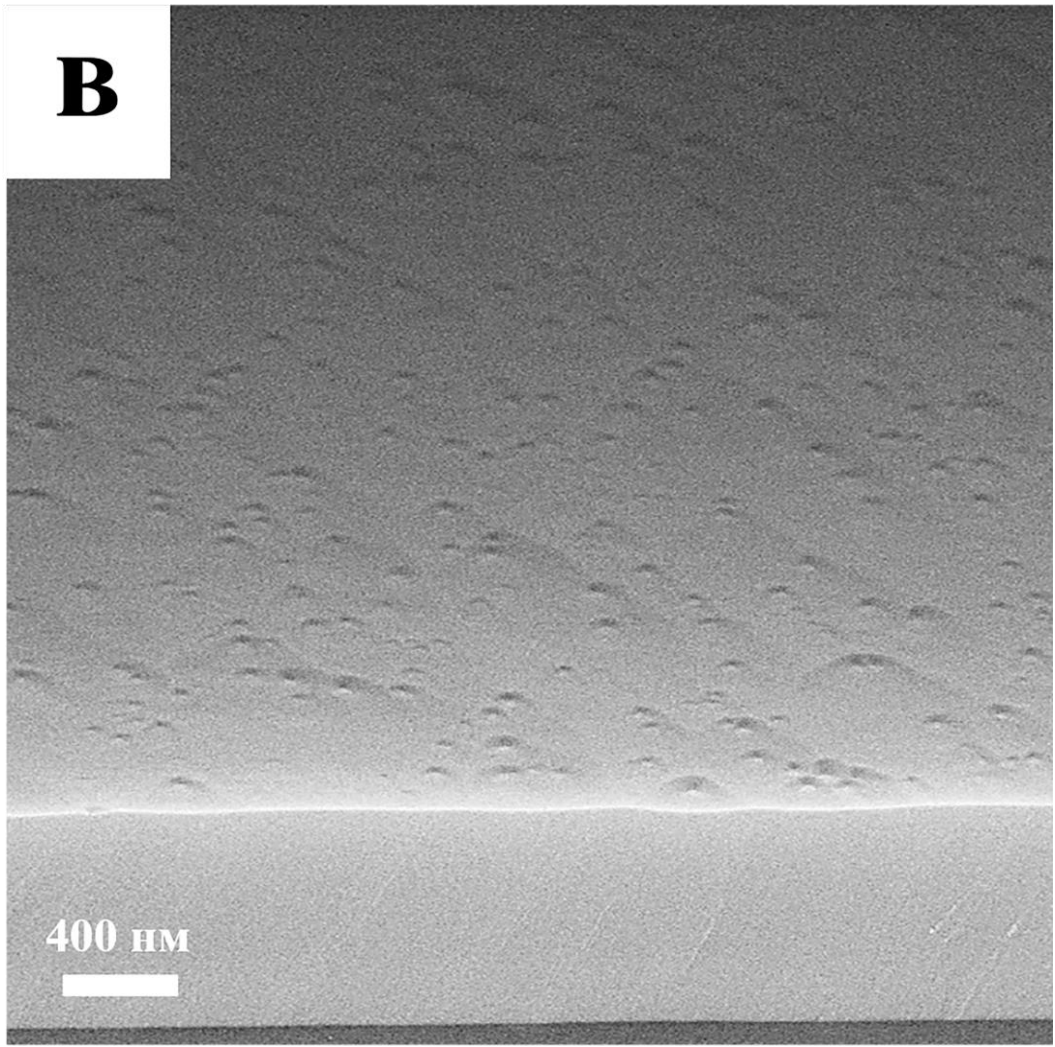

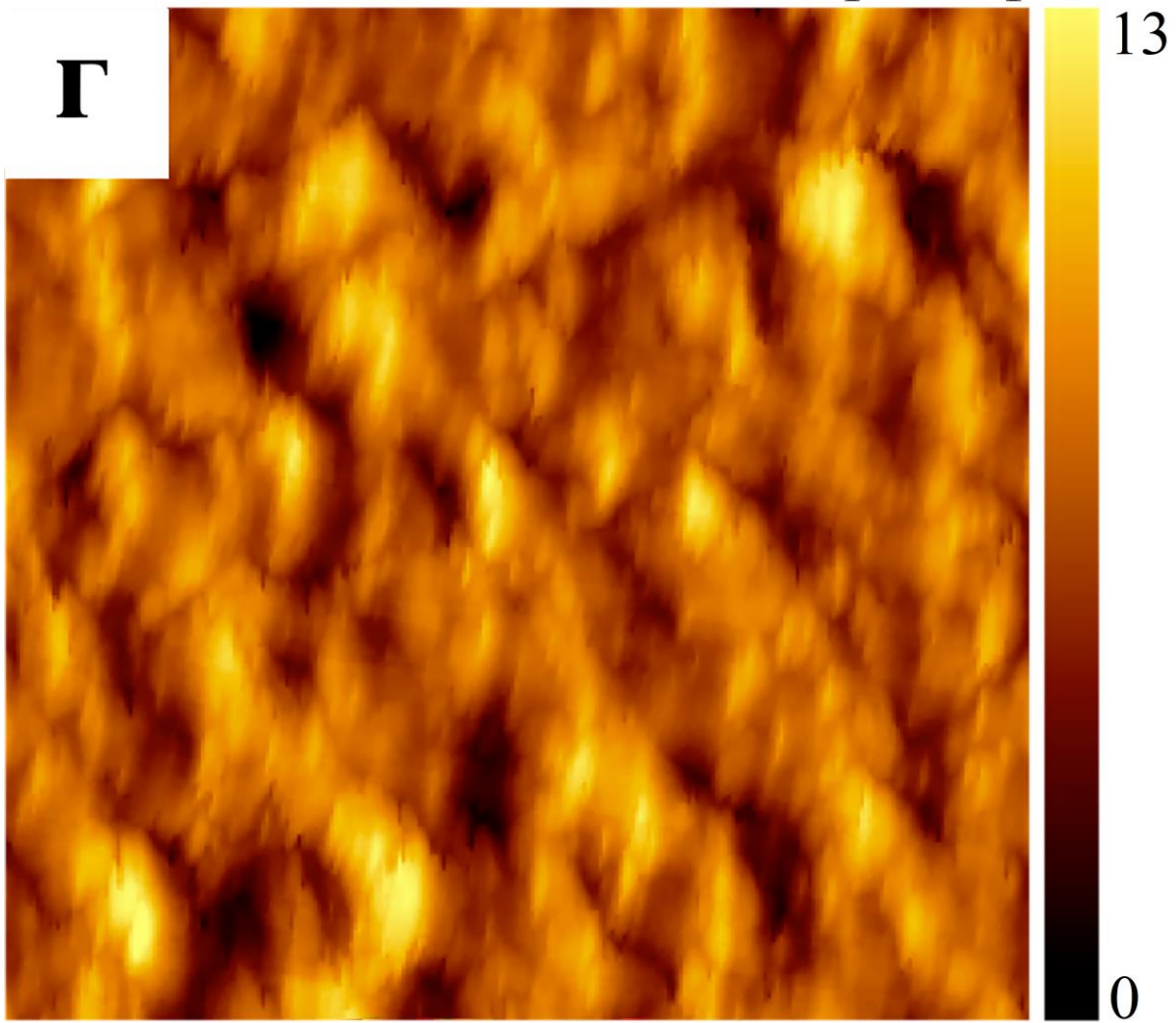
(г)

(д)

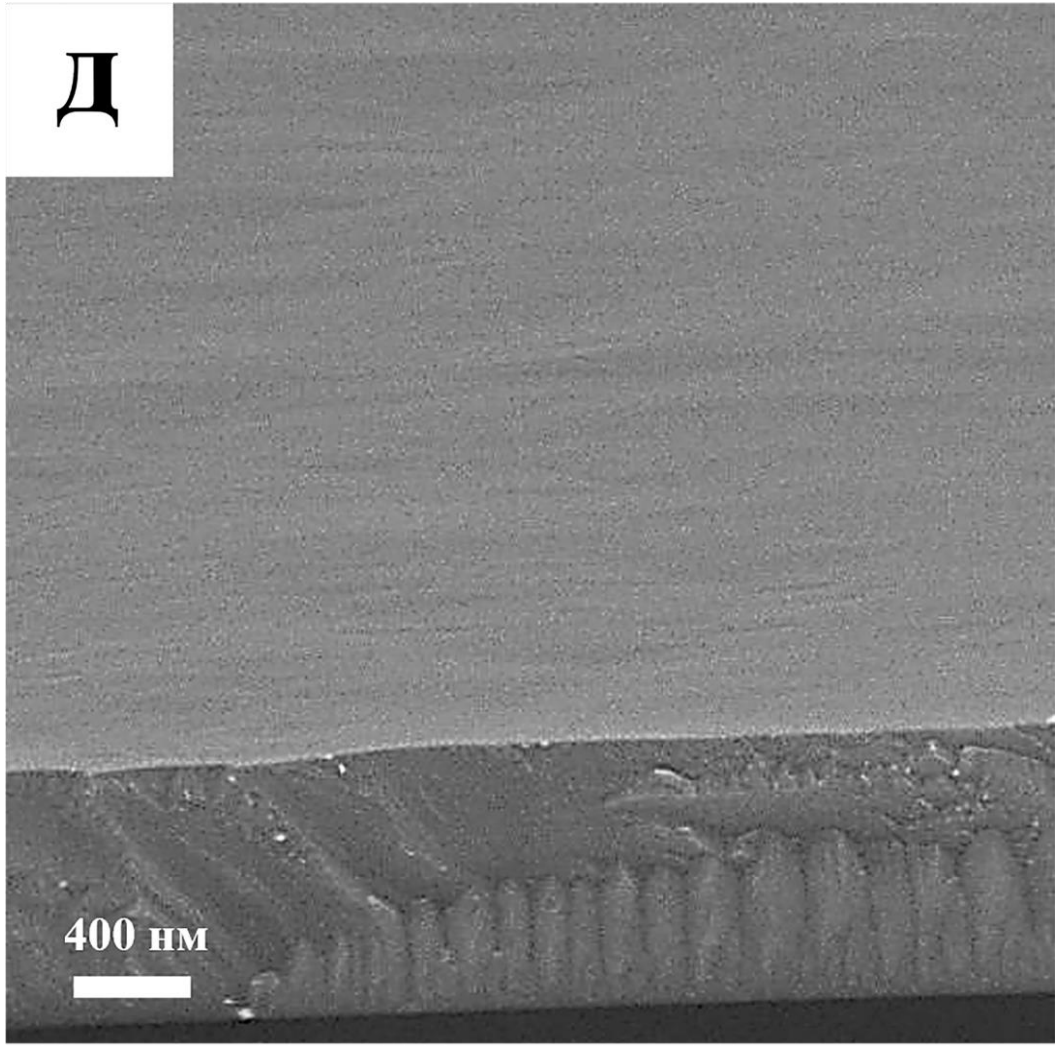

Рис. 1. Лендяшова. Поверхность.

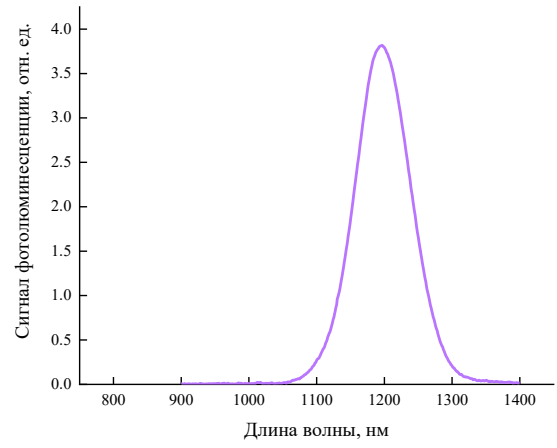

Рис. 2. Лендяшова. Поверхность.

**Адрес для переписки**

**Вера Вадимовна Лендяшова, - автор для связи**

**лаборант,**

194021, Санкт-Петербург,

*e-mail: erilerican@gmail.com*

Телефон: +7(921)7760760

**Игорь Владимирович Илькив, с.н.с.**

194021, Санкт-Петербург,

*e-mail: fiskerr@ymail.com*

Телефон: +7(904)6082025